\documentclass[a4paper,11pt]{article}
\usepackage{pos}

\usepackage{tcolorbox}
\usepackage{colortbl}

\title{Beyond a single elliptic curve}

\author{Hildegard M\"uller}
\author*{Stefan Weinzierl}

\affiliation{PRISMA Cluster of Excellence, Institut f{\"u}r Physik, \\
Johannes Gutenberg-Universit{\"a}t Mainz, \\
D - 55099 Mainz, Germany}

\emailAdd{hildemue@students.uni-mainz.de}
\emailAdd{weinzierl@uni-mainz.de}

\abstract{
In this talk we discuss the interplay of two elliptic curves, which occur in different sub-sectors of Feynman integrals.
We analyse a particular Feynman integral depending on two elliptic curves and derive an associated differential equation
in $\varepsilon$-form. We discuss the mixed entries of the differential equation, which depend on both elliptic curves.
}

\FullConference{%
  Loops and Legs in Quantum Field Theory - LL2022,\\
  25-30 April, 2022\\
  Ettal, Germany
}

\newcommand{\bq}{\begin{eqnarray}}
\newcommand{\eq}{\end{eqnarray}}
\newcommand{\eps}{\varepsilon}

\newcommand{\curveone}{(a)}
\newcommand{\curvetwo}{(b)}

\begin{document}
\maketitle


\section{Introduction}

The method of differential equations is a popular method to compute Feynman integrals \cite{Kotikov:1990kg,Kotikov:1991pm,Remiddi:1997ny,Gehrmann:1999as}.
We may systematically derive a differential equation for any Feynman integral
with the help of programs like {\tt FIRE} \cite{Smirnov:2008iw,Smirnov:2019qkx},
{\tt Reduze} \cite{Studerus:2009ye,vonManteuffel:2012np} and
{\tt Kira} \cite{Maierhoefer:2017hyi,Klappert:2020nbg}.
If we denote the vector of master integrals by
$I = (I_1,...,I_{N_F})^T$,
this gives us
\bq
 d I & = & A\left(\eps,x\right) I,
\eq
where the matrix-valued one-form $A$ depends on the dimensional regularisation parameter $\eps$ and the kinematic variables $x$.
Table~\ref{table_notation} summarises the notation used in this text.
The next and non-trivial step is to transform the differential equation into a particular nice form ($\eps$-form) \cite{Henn:2013pwa}: By a redefinition of the master integrals one tries to reach
\bq
 d I & = & \eps A\left(x\right) I
 \;\;\;\;\;\;
 \mbox{with}
 \;\;\;\;\;\;
 A \; = \; 
 \sum\limits_{j=1}^{N_L} \; C_j \; \omega_j,
\eq
where the $C_j$'s are $N_F \times N_F$-matrices, whose entries are numbers,
the only dependence on $\eps$ is given by the explicit prefactor
and the differential one-forms $\omega_j$ are closed and have only simple poles (see for example ref.~\cite{Weinzierl:2022eaz} for an introduction).
If such a transformation can be found (and appropriate boundary conditions are known), 
the differential equation can be solved order-by-order in $\eps$ in terms of iterated integrals \cite{Chen}.
\begin{table}
\begin{center}
\begin{tabular}{|llll|}
\hline
$I$ & $= (I_1, ..., I_{N_F})$ & Master integrals & \\
$N_F$ & $= N_{\mathrm{Fibre}}$ & Number of master integrals & \\
 \hline
$x$ & $=(x_1, ..., x_{N_B})$ & Kinematic variables &  \\ 
$N_B$ & $= N_{\mathrm{Base}}$ & Number of kinematic variables & \\
 \hline
$\omega$ & $=(\omega_1, ..., \omega_{N_L})$ & Differential one-forms/letters & \\ 
$N_L$ & $= N_{\mathrm{Letters}}$ & Number of letters & \\
\hline
\end{tabular}
\end{center}
\caption{
The notation used in the text.
}
\label{table_notation}
\end{table}
In this way the problem of computing Feynman integrals is reduced to finding an appropriate transformation 
for the differential equation.

We may now ask if for any family of Feynman integrals such a transformation can be found, and if yes,
what are the differential one-forms $\omega_j$ appearing in the differential equation?
Supporting evidence for the first part of the question is given by
many examples of Feynman integrals, which evaluate to multiple polylogarithms
and a few known integrals depending on a single elliptic curve \cite{Adams:2018yfj,Bogner:2019lfa}.
In this talk we report on a Feynman integral depending on two elliptic curves \cite{Muller:2022gec}.
This goes beyond the previously known cases.

Concerning the second part of the question:
The differential one-forms we know so far include
dlog-forms with possibly algebraic arguments, modular forms times $d\tau$ \cite{Adams:2017ejb}
and differential one-forms related to the Kronecker function \cite{Zagier:1991,Brown:2011,Broedel:2018qkq,Duhr:2019rrs,Weinzierl:2020fyx}.
The latter define elliptic polylogarithms \cite{Broedel:2017kkb}.
The particular Feynman integral in this talk depends on two elliptic curves and we expect to find one-forms $\omega_j$, which
go beyond the currently known ones.


\section{The Feynman integral}

\begin{figure}
\begin{center}
\includegraphics[scale=1.0]{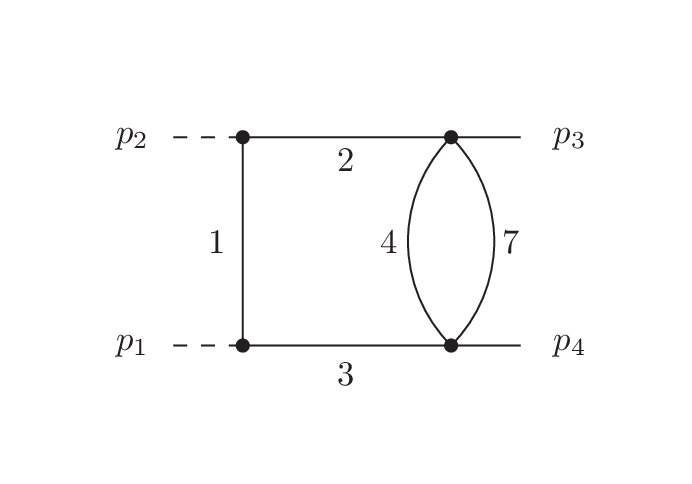}
\end{center}
\caption{
The graph for the Feynman integral.
Solid lines correspond to particles of mass $m$, dashed lines to massless particles.
}
\label{fig_sector_79}
\end{figure}

Fig.~\ref{fig_sector_79} shows the graph of the Feynman integral of interest.
This Feynman integral is a sub-topology of the double-box integral with an internal top-loop relevant to
top-pair production at the LHC \cite{Adams:2018bsn,Adams:2018kez}.
The numbering of the propagators follows the earlier publications.
We define the Mandelstam variables by
\bq
 s = \left(p_1+p_2\right)^2,
 & &
 t = \left(p_2+p_3\right)^2.
\eq
The Feynman integral depends on two dimensionless kinematic variables, which we may take originally as
\bq
 \frac{s}{m^2}, & & \frac{t}{m^2}.
\eq
There are $12$ master integrals in this family of Feynman integrals.
\begin{figure}
\begin{center}
\includegraphics[scale=0.4]{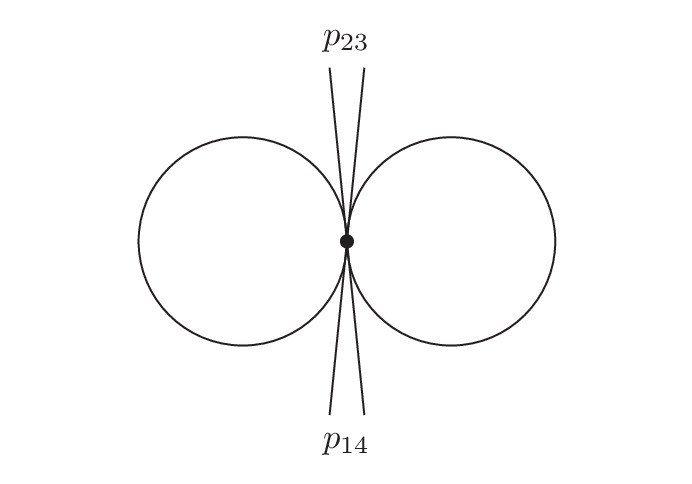}
\includegraphics[scale=0.4]{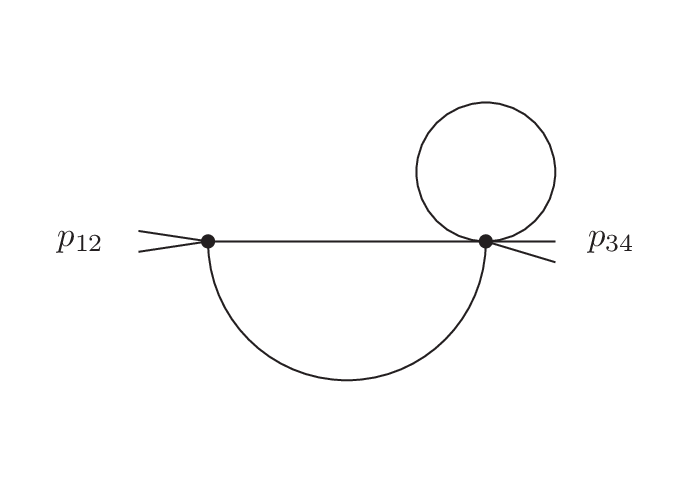}
\includegraphics[scale=0.4]{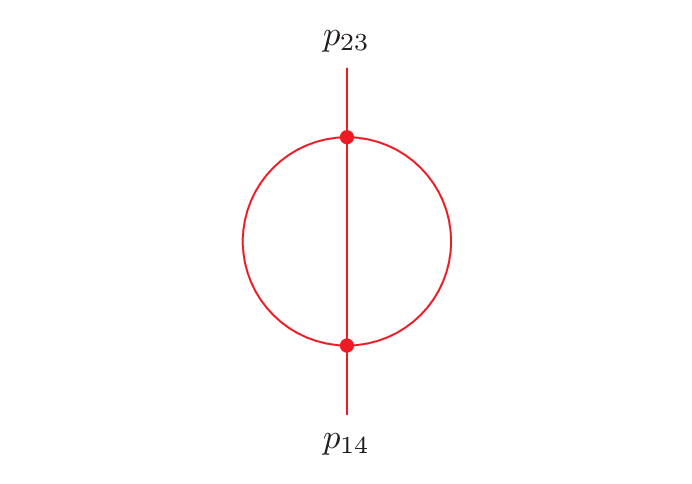}
\includegraphics[scale=0.4]{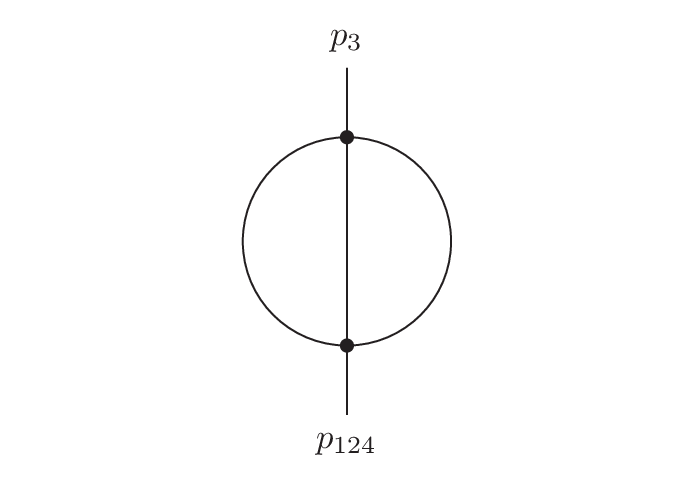}
\\
\includegraphics[scale=0.4]{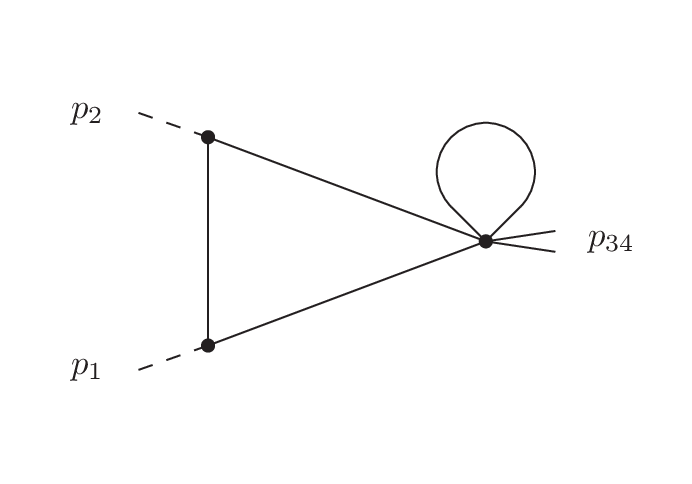}
\includegraphics[scale=0.4]{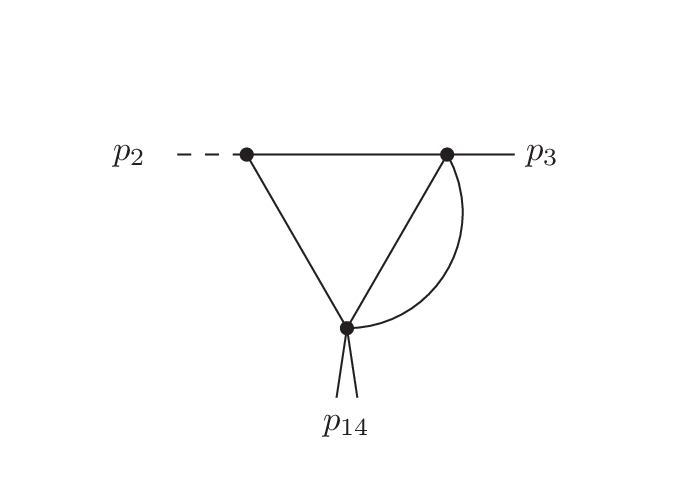}
\includegraphics[scale=0.4]{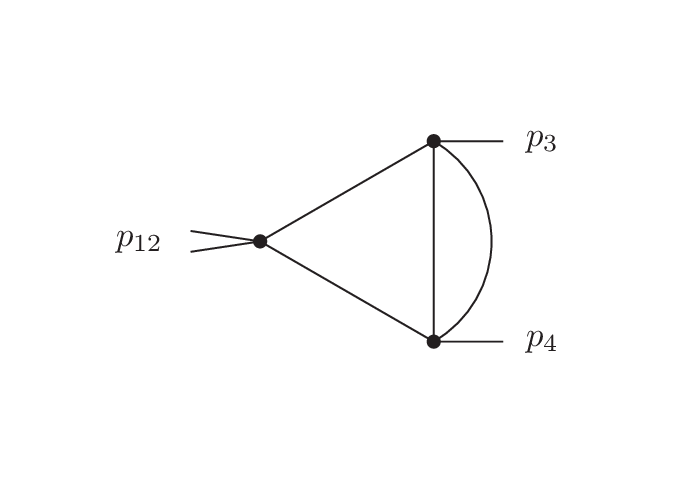}
\includegraphics[scale=0.4]{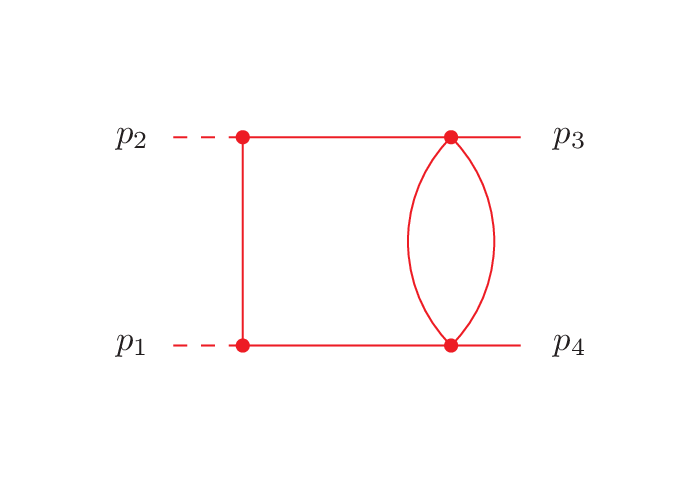}
\end{center}
\caption{
The master topologies.
There are $8$ master topologies.
A master topology may contain several master integrals.
The topologies corresponding to an elliptic curve are shown in red.
}
\label{fig_master_topologies}
\end{figure}
We may group the master integrals into master topologies, as shown in fig.~\ref{fig_master_topologies}.
There are eight master topologies, five of them have just one master integral,
two master topologies have two master integrals and one topology (the top topology)
has three master integrals.
We denote the Feynman integrals by $I_{\nu_1 \nu_2 \nu_3  \nu_4 \nu_5 \nu_6 \nu_7 \nu_8 \nu_9}(D)$, where $D$ denotes the number of space-time
dimensions and $\nu_j$ the power of the propagator $j$.

\section{The elliptic curves}

Two master topologies are associated with an elliptic curve. These two master topologies are shown in fig.~\ref{fig_master_topologies} 
in red.
These are the sunrise topology and the top topology.
We find the elliptic curves from the maximal cut of the master topology. We have
\bq
\lefteqn{
 \mathrm{MaxCut} \; I_{100100100}\left(2\right)
 \approx } & &
 \\
 & &
 \frac{m^2}{\pi^2}
 \int\limits_{\mathcal C} 
 \frac{du}{\sqrt{\left(u -t \right) \left(u - t + 4 m^2 \right) \left(u^2 + 2 m^2 u - 4 m^2 t + m^4\right)}},
 \nonumber \\
\lefteqn{
 \mathrm{MaxCut} \; I_{111200100}\left(4\right)
 \approx } & &
 \nonumber \\
 & &
 \frac{m^4}{4 \pi^3 s}
 \int\limits_{\mathcal C} 
 \frac{du}{\sqrt{\left(u -t \right) \left(u - t + 4 m^2 \right) \left(u^2 + 2 m^2 u - 4 m^2 t + m^4 - \frac{4m^2\left(m^2-t\right)^2}{s} \right)}}.
 \nonumber
\eq
The square roots in the denominators define the elliptic curves:
\bq
\mbox{Curve $\curveone$}: & &
 v^2 = \left(u -t \right) \left(u - t + 4 m^2 \right) \left(u^2 + 2 m^2 u - 4 m^2 t + m^4\right),
 \\
\mbox{Curve $\curvetwo$}: & &
 v^2 = \left(u -t \right) \left(u - t + 4 m^2 \right) \left(u^2 + 2 m^2 u - 4 m^2 t + m^4 - \frac{4m^2\left(m^2-t\right)^2}{s} \right).
 \nonumber 
\eq
For generic $(s,t)$, the two curves are neither isomorphic nor isogenic.
On the hypersurface $s=\infty$ the two curves are identical.

We choose two independent periods $\psi_1^{\curveone}$, $\psi_2^{\curveone}$ for curve $\curveone$.
The modular parameter $\tau^{\curveone}$ is then given by
$\tau^{\curveone} = \psi_2^{\curveone}/\psi_1^{\curveone}$.
We do the same for curve $\curvetwo$ and define the modular parameter $\tau^{\curvetwo}$ by
$\tau^{\curvetwo} = \psi_2^{\curvetwo}/\psi_1^{\curvetwo}$.
We may use $(\tau^{\curveone},\tau^{\curvetwo})$ instead of $(\frac{s}{m^2},\frac{t}{m^2})$
as kinematic variables.

\section{The differential equation}

The starting point is a differential equation, which is linear in $\eps$ and where the $\eps^0$-part 
is strictly lower triangular \cite{Adams:2018bsn,Adams:2018kez}:
\bq
 d I^{\mathrm{pre}} & = & \left[ A^{(0)}\left(x\right) + \eps A^{(1)}\left(x\right) \right] I^{\mathrm{pre}}.
\eq
The non-zero entries of $A^{(0)}$ are
\bq
 A^{(0)}_{11,10},
 \;
 A^{(0)}_{12,3},
 \;
 A^{(0)}_{12,5},
 \;
 A^{(0)}_{12,6},
 \;
 A^{(0)}_{12,7},
 \;
 A^{(0)}_{12,8},
 \;
 A^{(0)}_{12,10},
 \;
 A^{(0)}_{12,11}.
\eq
We first redefine the master integrals to put the differential equation into an $\eps$-form.
As $A^{(0)}$ is strictly lower triangular, this can be done by integration.
As an example we have for the eleventh master integral
\bq
 I_{11} \; = \; I_{11}^{\mathrm{pre}} + F_{11,10} I_{10}^{\mathrm{pre}},
 & &
 F_{11,10} \; = \; - \int A_{11,10}^{(0)}.
\eq
The exact definition of all master integrals can be found in \cite{Muller:2022gec}.
After the redefinition, the differential equation is in $\eps$-form:
\bq
 d I & = & \eps A\left(x\right) I.
\eq
The non-zero entries of $A$ are
\bq
\lefteqn{
 A
 = } & \\
 & &
 \left(\begin{array}{cccccccccccc}
 0 & 0 & 0 & 0 & 0 & 0 & 0 & 0 & 0 & 0 & 0 & 0 \\
 A_{2,1} & A_{2,2} & 0 & 0 & 0 & 0 & 0 & 0 & 0 & 0 & 0 & 0 \\
 \cellcolor{yellow} 0 & \cellcolor{yellow} 0 & \cellcolor{yellow} A_{3,3} & \cellcolor{yellow} A_{3,4} & 0 & 0 & 0 & 0 & 0 & 0 & 0 & 0 \\
 \cellcolor{yellow} A_{4,1} & \cellcolor{yellow} 0 & \cellcolor{yellow} A_{4,3} & \cellcolor{yellow} A_{4,4} & 0 & 0 & 0 & 0 & 0 & 0 & 0 & 0 \\
 0 & 0 & \cellcolor{yellow} 0 & \cellcolor{yellow} 0 & 0 & 0 & 0 & 0 & 0 & 0 & 0 & 0 \\
 0 & A_{6,2} & \cellcolor{yellow} 0 & \cellcolor{yellow} 0 & 0 & 0 & 0 & 0 & 0 & 0 & 0 & 0 \\
 0 & 0 & \cellcolor{yellow} A_{7,3} & \cellcolor{yellow} 0 & 0 & 0 & 0 & 0 & 0 & 0 & 0 & 0 \\
 0 & A_{8,2} & \cellcolor{yellow} 0 & \cellcolor{yellow} 0 & A_{8,5} & 0 & 0 & A_{8,8} & A_{8,9} & 0 & 0 & 0 \\
 0 & 0 & \cellcolor{yellow} 0 & \cellcolor{yellow} 0 & A_{9,5} & 0 & 0 & A_{9,8} & A_{9,9} & 0 & 0 & 0 \\
 \cellcolor{red} 0 & \cellcolor{red} 0 & \cellcolor{orange} A_{10,3} & \cellcolor{orange} 0 & \cellcolor{red} A_{10,5} & \cellcolor{red} A_{10,6} & \cellcolor{red} A_{10,7} & \cellcolor{red} A_{10,8} & \cellcolor{red} 0 & \cellcolor{red} A_{10,10} & \cellcolor{red} A_{10,11} & \cellcolor{red} A_{10,12} \\
 \cellcolor{red} 0 & \cellcolor{red} A_{11,2} & \cellcolor{orange} A_{11,3} & \cellcolor{orange} 0 & \cellcolor{red} A_{11,5} & \cellcolor{red} A_{11,6} & \cellcolor{red} A_{11,7} & \cellcolor{red} A_{11,8} & \cellcolor{red} A_{11,9} & \cellcolor{red} A_{11,10} & \cellcolor{red} A_{11,11} & \cellcolor{red} A_{11,12} \\
 \cellcolor{red} 0 & \cellcolor{red} A_{12,2} & \cellcolor{orange} A_{12,3} & \cellcolor{orange} A_{12,4} & \cellcolor{red} A_{12,5} & \cellcolor{red} A_{12,6} & \cellcolor{red} A_{12,7} & \cellcolor{red} A_{12,8} & \cellcolor{red} A_{12,9} & \cellcolor{red} A_{12,10} & \cellcolor{red} A_{12,11} & \cellcolor{red} A_{12,12} \\
 \end{array} \right).
 \nonumber
\eq
The colour coding is as follows:
Entries not highlighted by any colour are dlog-forms.
They only depend on $\frac{s}{m^2}$ (or $x$ defined by $\frac{s}{m^2}=-\frac{(1-x)^2}{x}$):
\bq
\label{dlog_forms}
 \frac{dx}{x},
 \;\;\;\;\;\;
 \frac{dx}{x-1},
 \;\;\;\;\;\;
 \frac{dx}{x+1}.
\eq
Entries highlighted in yellow are related to curve $\curveone$, but independent of curve $\curvetwo$.
They only depend on $\frac{t}{m^2}$ (or $\tau^{\curveone}$).
These entries are of the form 
\bq
 f_k(\tau^{\curveone}) \; 2\pi i d\tau^{\curveone},
\eq
where $f_k(\tau^{\curveone})$ is a modular form of $\Gamma_1(6)$.

Entries highlighted in red are related to curve $\curvetwo$, but independent of curve $\curveone$.
They depend on two kinematic variables $(\frac{s}{m^2},\frac{t}{m^2})$ or $(z^{\curvetwo},\tau^{\curveone})$.
These entries are of the form
\bq
\label{M_1_2_forms}
 f_k(\tau^{\curvetwo}) \; 2\pi i d\tau^{\curvetwo},
 & &
 \omega^{\mathrm{Kronecker}}_{k}\left(a z^{\curvetwo}+b,\tau^{\curvetwo}\right),
\eq
where $f_k$ is again a modular form (for the case at hand either of $\mathrm{SL}_2({\mathbb Z})$ or $\Gamma_0(2)$)
and
\bq
 \omega^{\mathrm{Kronecker}}_{k}
 & = &
 \left(2\pi i\right)^{2-k}
 \left[
  g^{(k-1)}\left( z, \tau\right) dz + \left(k-1\right) g^{(k)}\left( z, \tau\right) \frac{d\tau}{2\pi i}
 \right].
\eq
The coefficients $g^{(k)}(z, \tau)$ of the Kronecker function are defined as follows:
We first define the first Jacobi theta function $\theta_1(z,\bar{q})$ and the Kronecker function $F(z,\alpha,\tau)$ by
\bq
\theta_1\left(z,\bar{q}\right) 
 & = &
 -i \sum\limits_{n=-\infty}^\infty \left(-1\right)^n \bar{q}^{\frac{1}{2}\left(n+\frac{1}{2}\right)^2} e^{i \pi \left(2n+1\right)z},
 \nonumber \\
 F\left(z,\alpha,\tau\right)
 & = &
 \theta_1'\left(0,\bar{q}\right) \frac{\theta_1\left(z+\alpha, \bar{q}\right)}{\theta_1\left(z, \bar{q}\right)\theta_1\left(\alpha, \bar{q}\right)}.
\eq
As usual, we denote $\bar{q}=e^{2\pi i \tau}$ and
$\theta_1'(z,\bar{q})$ denotes the derivative with respect to $z$.
The coefficients $g^{(k)}(z, \tau)$ are then obtained from the expansion of 
the Kronecker function $F(z,\alpha,\tau)$ in $\alpha$:
\bq
 F\left(z,\alpha,\tau\right)
 & = &
 \frac{1}{\alpha} \sum\limits_{k=0}^\infty g^{(k)}\left(z,\tau\right) \alpha^k.
\eq
For a more detailed discussion of the functions $g^{(k)}(z, \tau)$ we refer to ~\cite{Weinzierl:2022eaz}
and the original references \cite{Zagier:1991,Brown:2011,Broedel:2018qkq,Duhr:2019rrs,Weinzierl:2020fyx}.
We may view the pair $(z^{\curvetwo},\tau^{\curveone})$ as coordinates on the moduli space ${\mathcal M}_{1,2}$ 
(the moduli space of a genus one curve with two marked points).
The variable $z^{\curvetwo}$ corresponds to a marked point on curve $\curvetwo$.
We have
\bq
 \omega^{\mathrm{Kronecker}}_{1}
 & = &
 2 \pi i dz^{\curvetwo}
\eq
and we may get $z^{\curvetwo}$ from integrating an entry of modular weight $1$ with respect to curve $\curvetwo$.
There are two entries of modular weight $1$ (i.e. $A_{10,11}$ and $A_{11,12}$). Picking the first one we obtain
\bq
 z^{\curvetwo} & = & \alpha + \beta \int A_{10,11}
\eq
up to two constants $\alpha$ and $\beta$.
As a side-remark let us mention that
the dlog-forms appearing in eq.~(\ref{dlog_forms}) can be written as a linear combination of the forms appearing in eq.~(\ref{M_1_2_forms}).

The entries highlighted in orange depend on both elliptic curves and are the most interesting ones.
The non-zero entries are 
$A_{10,3}$, $A_{11,3}$, $A_{12,3}$ and $A_{12,4}$.
We call these the ``mixed entries''.
For curve $\curveone$ we have the coordinate $\tau^{\curveone}$, 
for curve $\curvetwo$ we have the two coordinates $(z^{\curvetwo},\tau^{\curvetwo})$.
However, the Feynman integral depends only on two kinematic variables.
We may therefore express $\tau^{\curveone}$ as a function of $(z^{\curvetwo},\tau^{\curvetwo})$ 
or
$z^{\curvetwo}$ as a function of $(\tau^{\curveone},\tau^{\curvetwo})$.
Let us take $(\tau^{\curveone},\tau^{\curvetwo})$ as our main variables.
We may write any mixed entry as
\bq
 A_{i,j} & = & 
 A_{i,j}^{\curveone} \; 2 \pi i d\tau^{\curveone}
 +
 A_{i,j}^{\curvetwo} \; 2 \pi i d\tau^{\curvetwo}.
\eq
Integrability and information from limits allow us to fix these entries.
From integrability we obtain for example
\bq
 A_{12,4}^{\curvetwo}
 & = &
 i \frac{H_4\left(z^{\curvetwo},\tau^{\curvetwo}\right)}{g_3\left(\tau^{\curveone}\right)} 
 \frac{\partial z^{\curvetwo}}{\partial \tau^{\curveone}},
\eq
where $g_3$ is a modular form of modular weight $3$ with respect to curve $\curveone$ and
$H_4$ transforms quasi-modular of modular weight $4$ with respect to curve $\curvetwo$.
The exact definition of $g_3$ and $H_4$ can be found in \cite{Muller:2022gec}.
Let us introduce local primitives $\Omega_{i,j}$ such that
\bq
\label{def_primitive}
 A_{i,j} & = & d \Omega_{i,j}.
\eq
From the limit $\tau^{\curvetwo} \rightarrow i \infty$ one finds then for example
\bq
 \Omega_{12,4} 
 & = &
 \frac{1}{4} \ln \bar{q}^{\curveone}
 +
 2 \pi i
 \int d\tau^{\curvetwo}
 A_{12,4}^{\curvetwo}.
\eq
The full entry $A_{12,4}$ is then given by eq.~(\ref{def_primitive}).

Let us set
\bq
 \bar{q}^{\curveone} \; = \; e^{2\pi i \tau^{\curveone}},
 & &
 \bar{q}^{\curvetwo} \; = \; e^{2\pi i \tau^{\curvetwo}}.
\eq
For numerical evaluations 
the expansions in $\bar{q}^{\curveone}$ and $\bar{q}^{\curvetwo}$ are useful.
One finds for example
\bq
 \Omega_{12,4}
 & = &
 \frac{1}{4} \ln\bar{q}^{\curveone} - \frac{1}{2}\ln\left(\bar{q}^{\curvetwo}-\bar{q}^{\curveone}\right)
 \nonumber \\
 & &
 + \left[ 5 - 27 \bar{q}^{\curveone} + 53 \left(\bar{q}^{\curveone}\right)^2  + 552 \left(\bar{q}^{\curveone}\right)^3 + \dots \right] \bar{q}^{\curvetwo} 
 \nonumber \\
 & &
 - \frac{1}{2} \left[ -43 - 328 \bar{q}^{\curveone} + 11043 \left(\bar{q}^{\curveone}\right)^2  + \dots \right] \left(\bar{q}^{\curvetwo}\right)^2 
 \nonumber \\
 & &
 + \frac{1}{3} \left[ -526 + 20790 \bar{q}^{\curveone} + \dots \right] \left(\bar{q}^{\curvetwo}\right)^3 + \dots 
\eq
and the one for $A_{12,4}$ can again be obtained by differentiation.

\section{Conclusions}

In this talk we discussed a two-loop Feynman integral with four external legs and one internal mass,
depending on two kinematic variables.
This Feynman integral has two elliptic curves associated to it: One elliptic curve 
is associated to the maximal cut of the top sector, the second elliptic curve 
is associated to the sunrise sub-topology.
Our main results are threefold:
We first showed that the differential equation can be transformed to an $\eps$-form.
This result supports the conjecture that an $\eps$-form can be reached for any Feynman integral.

We then studied the entries of the differential equation, and here in particular the ones
giving the derivatives of the three master integrals in the top sector.
We found that most of these entries depend only on curve $\curvetwo$, but not on curve $\curveone$.
These entries can be expressed in terms of differential one-forms already encountered in other Feynman integrals.
This shows the universality of these differential one-forms.

Finally, we studied the entries which depend on both elliptic curves. 
We obtained a natural representation of the mixed entries in terms of the variables
$(\tau^{\curveone},\tau^{\curvetwo},z^{\curvetwo})$,
which makes the modular transformation properties with respect to the two elliptic curves transparent.

We point out that the $\eps$-form of the differential equation gives us the information on what possibly might appear
at any given order in the dimensional regularisation parameter $\eps$.
If we just look at the lowest non-vanishing order for each master integral, not all letters may appear there.
Of course it is desirable to disentangle the dependence on multiple elliptic curves as much as possible, as it was done for
the massless two-loop pentabox in four space-time dimensions in \cite{Bourjaily:2022tep}.


{\footnotesize
\bibliography{/home/stefanw/notes/biblio}
\bibliographystyle{/home/stefanw/latex-style/h-physrev5}
}

\end{document}